\documentclass[twocolumn,prb,showpacs]{revtex4}
\usepackage{graphicx}
\usepackage{bm}

\begin{document}

\title{Enhanced Spin Dependent Shot Noise in Magnetic Tunnel Barriers}

\author{Samir Garzon$^1$\footnote{sgarzon@physics.sc.edu}, Yuanzhen Chen$^{2}$, and Richard A. Webb$^{1}$}

\affiliation{$^1$Department of Physics and Astronomy and USC Nanocenter, University of South Carolina, Columbia, SC 29208, USA\\
$^2$Department of Physics and Astronomy, University of Pennsylvania,
Philadelphia, PA 19104, USA}

\begin{abstract}
We report the observation of enhanced spin dependent shot noise in
magnetic tunnel barriers, suggesting transport through localized
states within the barrier. This is supported by the existence of
negative magnetoresistance and structure in the differential
conductance curves. A simple model of tunneling through two
interacting localized states with spin dependent tunneling rates is
used to explain our observations.
\end{abstract}


\maketitle

\section{Introduction}\label{Introduction}

Measurement of fluctuations or noise for gaining deeper
understanding of the microscopic details of physical systems is not
new. Shot noise measurements have been successfully used to measure
the transport of fractional charge in two dimensional electron gas
(2DEG) Hall regime
systems~\cite{saminadayar:1997,de-picciotto:1997}, to study 2e
charge transport in superconductor-metal
interfaces~\cite{jehl:1999,jehl:2000,kozhevnikov:2000}, to detect
localized states in point contact experiments in
2DEG's~\cite{chen:2006PRB,safonov:136801}, and to probe transport
details of many other
systems~\cite{dieleman:1997,reznikov:1995,kumar:1996,vanderbrom:1999,li:1990,steinbach:1996,liu:1998}.
However, shot noise in magnetic systems has received much less
attention. It is only recently that some predictions of the
dependence of the shot noise on parameters such as the degree of
spin polarization have been
made~\cite{tserkovnyak:2001,brataas:2000,lamacraft:2004,zareyan:2005,belzig:2004,mishchenko:2003,sauret:2004,braun:075328,bulka:2000,bulka:1999,belzig:2005,elste:2006,cottet:2004}.
Corresponding shot noise experiments in magnetic systems are few,
and have been done in frequency regimes where 1/f noise
dominates~\cite{nowak:1999}. In this paper we report the observation
of enhanced spin dependent shot noise in magnetic tunnel
junctions~\cite{privcomm}. Enhancement of the shot noise above the
Poissonian limit in magnetic systems can occur via ``spin blockade"
due to the presence of a localized state, where a minority spin
electron (with lower tunneling rate) can block the transport of
majority spin electrons (with higher tunneling
rate)~\cite{braun:075328,bulka:2000,bulka:1999,belzig:2005,elste:2006,cottet:2004}.
In nonmagnetic tunnel barriers it is also possible to observe
super-Poissonian noise which has its origin in a similar type of
blocking behavior whenever two localized states within the barrier
are available for transport. An electron which tunnels into the
state with the lower tunneling rates can block transport through the
other localized state with higher tunneling rates, due to Coulomb
interaction (``charge blockade")~\cite{safonov:136801}. Since in
magnetic tunnel barriers both spin and charge blockade might occur,
we compare our measurements with the results from calculations using
both models. Our results are consistent with the model of two
interacting localized states with spin dependent tunneling rates,
implying that the observed super-Poissonian shot noise is due to
both charge and spin blockade.

The paper is organized as follows. Section two discusses enhanced
shot noise in nonmagnetic tunnel barriers due to two interacting
localized states~\cite{safonov:136801} and explains a classical
model that can be used to calculate the Fano factor for arbitrary
tunnel barriers~\cite{buttcomm}. Section three describes an
additional mechanism for enhanced shot noise, particular only to
magnetic tunnel junctions, and extends the model described in
section two to include spin dependent tunneling rates. Section four
describes the sample fabrication and experimental setup, while the
results of the measurements and analysis are presented in section
five. Finally, section six shows our conclusions.

\section{Shot Noise Enhancement in Nonmagnetic Tunnel Junctions}

Electron transport across a tunnel junction can be characterized by
a set of transmission coefficients $T_{n}$, associated with the
conducting channels across the junction. Within the
Landauer-B\"{u}ttiker formalism the total noise in a tunnel junction
(both thermal plus non-equilibrium excess noise) is given by

\begin{eqnarray}
\label{equ:normaljunction1} S{(f)}=&&\frac{2e^{2}}{\pi\hbar}~
[2k_{B}T\sum_{n}T_{n}^{2}+\nonumber\\
&&eV\coth{(\frac{eV}{2k_{B}T})}\sum_{n}T_{n}(1-T_{n})],
\end{eqnarray}

\noindent where $f$, $T$, and $V$ are the frequency, temperature,
and voltage bias~\cite{blanter:2000}. In the limit of $eV \gg
k_{B}T$, Eq. (\ref{equ:normaljunction1}) reduces to

\begin{equation}
\label{equ:normaljunction2} S(f) =
\frac{4e^{3}|V|}{h}\sum_{n}T_{n}(1-T_{n}),
\end{equation}

\noindent hence the Fano factor of such a junction is given by

\begin{equation}
\label{equ:fano1} F \equiv \frac{S(f)}{2eI}=
\frac{\sum_{n}T_{n}(1-T_{n})}{\sum_{n}T_{n}}.
\end{equation}

\begin{figure}[tb]
\begin{center}\leavevmode
\includegraphics[width=0.98\linewidth]{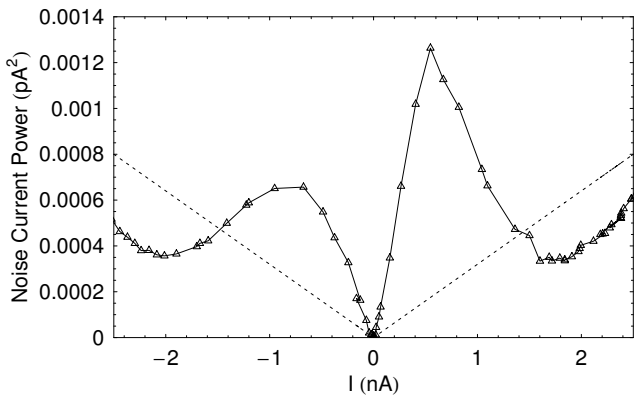}
\caption{Noise current power as a function of current in a 2DEG
point contact at 70 mK at a fixed gate voltage, together with the
calculated full shot noise (dashed
line).}\label{fig:chennoise}\end{center}\end{figure}

\noindent $F$ cannot exceed one, that is, a regular tunnel junction
exhibits only suppressed shot noise. This is equivalent to the
statement that the transport statistics of a system with multiple
uncorrelated sequential and/or parallel transport channels, each of
which can be described by a Poissonian probability distribution, is
exclusively sub-Poissonian. This was verified to a very good
precision in atomic size metallic tunnel
junctions~\cite{vanderbrom:1999}. It thus came as a surprise when
experimental observation of enhanced shot noise in a tunnel junction
(which requires super-Poissonian statistics) was first
reported~\cite{safonov:136801}. The enhancement was explained using
a model of interacting localized states inside the tunnel junction.
Depending on the occupation state of one localized site, tunneling
through the other site could be significantly modified due to the
strong Coulomb interaction between the two sites. Such a modulated
tunneling process could lead to largely enhanced shot noise,
specially in small area tunnel barriers 
where the interaction between localized states is strongest. Our
measurements in tunnel junctions formed by an electrostatic
potential in 2DEG systems have also shown shot noise enhancement due
to localized states~\cite{chen:2006PRB,chen:2006}. Shot noise in
these junctions is very sensitive to microscopic details such as the
distribution of localized states in real space, as well as their
energy landscape. Figure \ref{fig:chennoise} shows the shot noise as
a function of current for one of our 2DEG point contacts for a fixed
gate voltage. The shot noise can be either enhanced or suppressed by
changing the current bias through the point contact, and is in
general asymmetric with respect to the bias, reflecting the
asymmetry in the position of the localized state. In our
measurements, Fano factors between 0.3 and 14 have been observed on
25 barriers.

The probability distribution of tunneling events through a tunnel
barrier with an arbitrary number of either sequential or parallel,
and either interacting or independent transport channels can be
constructed within a simple classical model using single channel
tunneling events, each of which is characterized by a Poissonian
probability distribution with tunneling rate $\Gamma_i$. The
resulting total probability distribution $P(t)$ for an electron to
tunnel completely across the barrier at exactly a time t after the
previous tunneling event, can be either sub-Poissonian, Poissonian,
or super-Poissonian. For particles with charge $e$ tunneling between
two reservoirs with a general probability distribution $P(t)$, the
average current $I$ is given by

\begin{equation}
I=\frac{e}{\langle t\rangle} \label{equ:averageI}
\end{equation}

\noindent and the Fano factor by~\cite{davies:1992}

\begin{equation}
F=\frac{\langle t^2\rangle}{\langle t\rangle^2}-1,
\label{equ:Fanoall}
\end{equation}

\noindent where for a general function $g(t)$ the expectation value
$\langle g(t)\rangle$ is defined as $\langle
g(t)\rangle\equiv\int_0^\infty g(t) P(t) dt$.

For the particular case of tunneling through two interacting
localized states we assume that the tunneling rates associated with
each site are $\Gamma_{iL}$ and $\Gamma_{iR}$ ($i$ = 1, 2). We
further assume that due to Coulomb interaction once an electron hops
into one state the other state will become unavailable for other
electrons. The total probability distribution in this case is given
by

\begin{eqnarray}
P(t)=\int_{0}^{t}dt'e^{-\Gamma_{1L}t'}&&e^{-\Gamma_{2L}t'}(e^{-\Gamma_{2R}(t-t')}\Gamma_{2L}\Gamma_{2R}\nonumber\\
&&+e^{-\Gamma_{1R}(t-t')}\Gamma_{1L}\Gamma_{1R}). \label{equ:P2loc}
\end{eqnarray}

\noindent If the four tunneling rates are related by $\Gamma_{1L}$ =
$\alpha\Gamma_{1R}$ = $\beta\Gamma_{2L}$ = $\beta\gamma\Gamma_{2R}$,
then the Fano factor calculated from Eqs. \ref{equ:Fanoall} and
\ref{equ:P2loc} is

\begin{equation}
\label{equ:fano3} F =
\frac{2\beta^2\gamma^2+\beta(1+(\alpha-\gamma)^2)+2\alpha^2}
{\beta(1+\alpha+\gamma)^2}.
\end{equation}

\begin{figure}[tb]
\begin{center}\leavevmode
\includegraphics[width=0.98\linewidth]{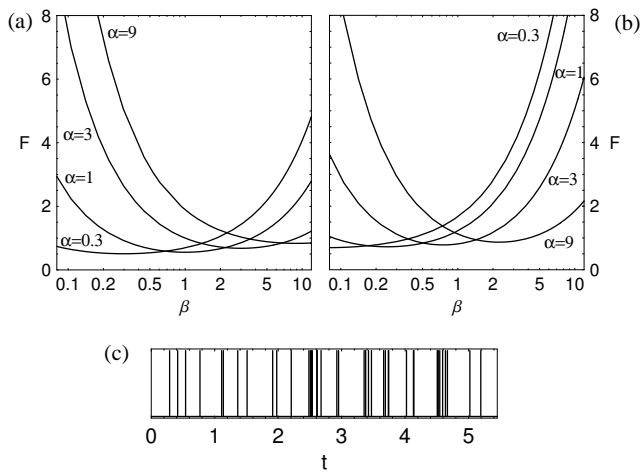}
\caption{Fano factor as a function of the asymmetry in tunneling
rates between two localized states $\beta$. Curves for different
values of the left-right asymmetry parameter of localized state 1,
$\alpha$, are shown for (a) $\gamma$=1 (no left-right asymmetry in
localized state 2) and (b) $\gamma$=4. (c) Simulation of tunneling
events as a function of time showing bunching of
electrons.}\label{fig:locfan}\end{center}\end{figure}

\noindent Parameters $\alpha$ and $\gamma$ describe the asymmetry
between left and right tunneling rates for each localized state,
while $\beta$ describes the asymmetry in tunneling rates of the two
localized states. Figure \ref{fig:locfan} shows $F$ as a function of
$\beta$ at different values of $\alpha$ and $\gamma$. The Fano
factor has a minimum value of 0.5 when $\beta$ =
$\frac{\alpha}{\gamma}$ and $\gamma$ = $1-\alpha$. On the other
hand, $F$ has no upper bound. In particular, $F$ can be much larger
than 1 when $\beta$ is far away from 1, that is, when electrons
dwell much longer in one site than in the other, effectively
blocking transport through the faster channel.
 Even tunneling
rates which are less than one order of magnitude different give
large values of $F$. Since the tunneling rate depends exponentially
on the distance between the localized state and the reservoir, even
a small asymmetry in the position of the localized state within the
barrier will give large differences in the left and right tunneling
rates (this is specially true whenever the localization length is
small). Therefore whenever two localized states are close enough so
that one channel can block transport through the other, $F>1$ will
be observed more often than $F<1$. We performed numerical
simulations to study the tunneling events in the time domain. A
typical result is given in Fig. \ref{fig:locfan}(c), where the
``bunching" pattern in the tunneling events, which leads to enhanced
shot noise, is clear. Unlike the ``bunching" behavior observed for
bosons where quantum statistics plays a
role~\cite{hanbury_brown:1956,morgan:1966,paul:1982,blanter:2000},
here it is purely due to Coulomb interaction. For magnetic tunnel
barriers, however, another mechanism can also lead to ``bunching"
effects. This will be discussed in the next section.


\section{Magnetic Tunnel Junctions}

In nonmagnetic tunnel barriers the tunneling rates for electrons
with any spin orientation are considered to be equal. However in
magnetic tunnel barriers this assumption is incorrect since the
density of states in each of the magnetic electrodes is spin
dependent. Therefore tunneling rates are larger for majority than
for minority spins. This provides a different mechanism for
generation of super-Poissonian statistics. Following the procedure
described in the previous section, we can calculate the probability
distribution for an electron to tunnel exactly at a time t after the
previous tunneling event through a tunnel barrier with one localized
state and spin dependent tunneling rates. It has the same form as
Eq. \ref{equ:P2loc}, replacing the indices 1 and 2 by $\uparrow$ and
$\downarrow$. This is not surprising since in both situations (i)
electrons tunnel into a localized state at two different rates
($\Gamma_{1L}$,$\Gamma_{2L }$ in the case of 2 localized states, or
$\Gamma_{\uparrow L}$,$\Gamma_{\downarrow L}$ for spin dependent
tunneling and a single localized state) and (ii) only one localized
state can be occupied at any time. We have assumed that an external
bias is applied in such a way that tunneling occurs from the left to
the right reservoir. Assuming that $\Gamma_{\uparrow i}=\gamma_i
(1+p)$ and $\Gamma_{\downarrow i}=\gamma_i (1-p)$ for $i=R,L$, with
$p$ the spin polarization (assumed to be the same for both magnetic
electrodes) and $\gamma_i$ the average tunneling rate for an
electron with a definite spin state~\cite{bulka:2000}, the Fano
factor is given by

\begin{equation}
F_P=\frac{4 \gamma_L^2\frac{1+p^2}{1-p^2}+\gamma_R^2}{(2
\gamma_L+\gamma_R)^2} \label{equ:spinblocP}
\end{equation}

\noindent for the parallel magnetization state (P), while for the
antiparallel state (AP) it is

\begin{equation}
F_{AP}=\frac{4 \gamma_L^2\frac{1+4p^2-p^4}{(1-p^2)^2}+\gamma_R^2}{(2
\gamma_L\frac{1+p^2}{1-p^2}+\gamma_R)^2}. \label{equ:spinblocAP}
\end{equation}

\begin{figure}[tb]
\begin{center}\leavevmode
\includegraphics[width=0.98\linewidth]{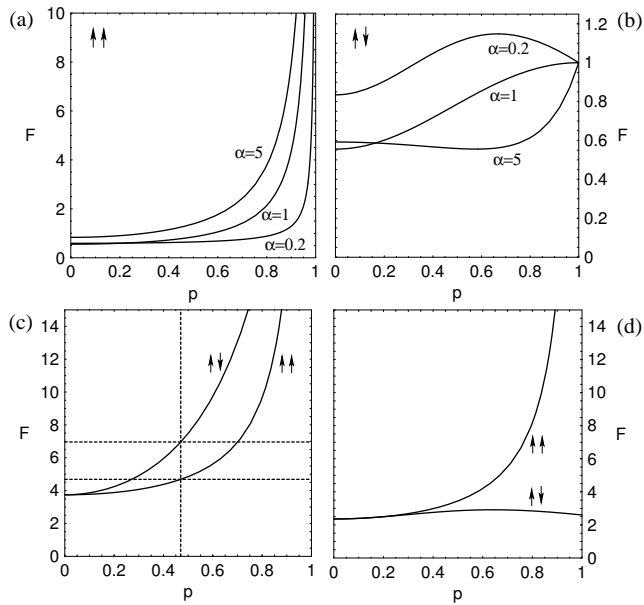}
\caption{Fano factor as a function of the spin polarization for (a)
parallel (P) and (b) antiparallel (AP) magnetization states in a
magnetic tunnel barrier with a single localized state. Curves with
different left-right asymmetry parameter $\alpha=\gamma_L/\gamma_R$
are shown. (c,d) Fano factor as a function of the spin polarization
for two localized states with parameters (c) $\alpha=0.05$,
$\beta=35$, $\gamma=0.15$ and (d) $\alpha=\beta=\gamma=5$ for both P
and AP states. }\label{fig:spinbloc}\end{center}\end{figure}

\noindent Eq. \ref{equ:spinblocP} agrees with the calculations of
Braun et. al~\cite{braun:075328} who only give a closed form for the
P state. For $p=0$ Eqs. \ref{equ:spinblocP} and \ref{equ:spinblocAP}
are equal and reduce to the case of nonmagnetic tunnel barriers,
with tunneling rate from the left $2 \gamma_L$ and tunneling rate to
the right $\gamma_R$. This result shows that, in contrast to what is
typically calculated for tunneling between nonmagnetic reservoirs
through a barrier with a localized state, the minimum Fano factor
$F=1/2$ occurs for an asymmetric position of the localized state,
when $\gamma_R=2\gamma_L$. Tunneling into a localized state is
proportional to the total density of states in the incoming
reservoir but tunneling out of a localized state is proportional
only to the density of states with spin $\overrightarrow{S}$, with
$\overrightarrow{S}$ the spin of the electron in the localized
state~\cite{spinflip}. Therefore, for a symmetric localized state
the left and right tunneling rates are different and give $F\neq
1/2$. Plots of the Fano factor as a function of the spin
polarization $p$ are shown in Fig. \ref{fig:spinbloc} for (a) the P
and (b) the AP states for different asymmetries between left and
right tunneling rates. For large spin polarization the Fano factor
in the P state can be much larger than 1, while in the AP state it
has a maximum of 1.25 for $p=1/\sqrt{3}$ and$\gamma_L\gg\gamma_R$.
This behavior has been termed ``spin
blockade"~\cite{braun:075328,bulka:2000,bulka:1999,belzig:2005,elste:2006,cottet:2004}.
It is important to note that both P states ($\uparrow\uparrow$ and
$\downarrow\downarrow$) have the same Fano factor $F_P$ and both AP
states ($\uparrow\downarrow$ and $\downarrow\uparrow$) have the same
Fano factor $F_{AP}$. Only large spin polarizations produce
significant shot noise enhancement.

It is then natural to consider the case of two interacting localized
states, as discussed in the previous section, but now for magnetic
tunnel barriers. This can be done in the same spirit as before,
including spin dependent tunneling rates from the left reservoir to
each of the localized states, and from the localized states to the
right reservoir. Assuming equal spin polarization for both
reservoirs, the set of 8 rates that describe the tunneling process
can be reduced to 5 parameters, $\Gamma_L$, $\alpha$, $\beta$,
$\gamma$, and p, which have been already defined in sections 2 and
3; however, the Fano factor is independent of $\Gamma_L$. In this
scenario the Fano factor of the AP state can be much larger than
1.25 (the maximum obtainable value for a single localized state).
Furthermore both possibilities $F_P>F_{AP}$ and $F_{AP}>F_P$ can
occur, as illustrated in Figures \ref{fig:spinbloc}(c,d).

\section{Experiment}

Samples are fabricated on Si wafers with 500 nm of thermally grown
oxide with a standard electron-beam lithography, thermal
evaporation, and liftoff technique. A bilayer of PMGI and PMMA is
used to create large controllable undercuts in the resist profile,
which allows the use of double angle evaporation techniques, and
therefore higher quality tunnel barriers that can be completely
deposited without breaking vacuum, which in our system is typically
in the 10$^{-7}$ Torr range. First, 15 nm of Co are thermally
evaporated at a low rate (typically 0.02 nm/s), after which a thin
(2-4 nm) layer of Al is thermally evaporated at 0.01 nm/s. Then, the
Al is oxidized at 800 to 2000 Torr-s, and the top 40 nm Co layer is
deposited at a higher rate. The two Co electrodes have different
geometry: the bottom electrode is typically 350 nm by 1.4 $\mu$m
while the top one is 120 nm by 1.4 $\mu$m. Devices with different
overlap areas between the two Co electrodes (typically between 0.005
nm$^2$ and 0.06 nm$^2$) are fabricated in order to achieve
resistances in the proper range. The resistance of the devices is
measured at room temperature, and samples with resistances lower
than 30k$\Omega$ and higher than 300k$\Omega$ are discarded. Within
a batch, most of the samples with similar Co overlap areas have
similar resistance values. Devices are cooled down to 8K, and the
resistance is measured again to ascertain if they have
insulator-type of behavior and the resistance has increased. If
metallic behavior is observed (i.e. the resistance decreases as the
temperature is lowered) the sample is discarded since this shows
that pinhole conduction is taking
place~\cite{mukhopadhyay:026601}~(Rowell's
criteria~\cite{rowell:1969}).

\begin{figure}[tb]
\begin{center}\leavevmode
\includegraphics[width=0.98\linewidth]{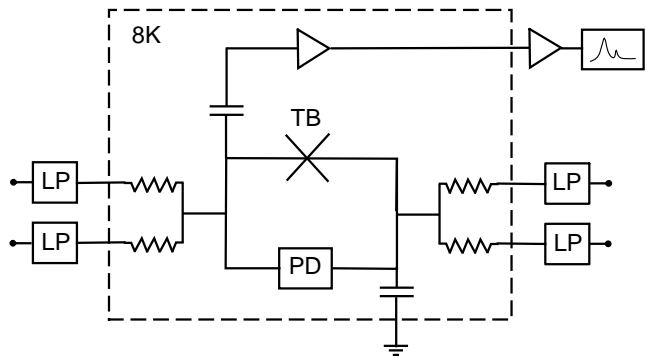}
\caption{Schematic of measurement configuration showing four lowpass
filtered leads with isolation resistors, a photodiode (PD) in
parallel with the tunnel barrier (TB), and the two stages of
amplification before the spectrum analyzer. Blocking capacitors are
used to separate the DC bias current from the 100kHz
noise.}\label{fig:diagram}\end{center}\end{figure}

Measurements are performed in a variable temperature cryostat
equipped with an axial 9 T magnet (Quantum Design) and a custom
built probe. For shot noise measurements the device is dc current
biased while the noise voltage across the tunnel barrier is measured
using a cryogenic preamplifier mounted less than 5 cm away from the
sample. A complementary room temperature stage is used for further
amplification after which a spectrum analyzer (HP89410A) samples the
voltage noise in a flat frequency band in the 100kHz range where the
effects of 1/f noise can be neglected. In situ calibration of the
measured voltage is done with a photodiode (Lasermate, PDT-A85A30)
wired in parallel to the tunnel barrier (see Fig.
\ref{fig:diagram}). A photocurrent with full shot noise is generated
in the photodiode by
 means of an LED,
while the voltage drop across the tunnel barrier is nulled by
applying an external current. This ensures that all of the current
dependent noise comes from the photodiode. A single calibrated
measurements consists of measuring the photodiode noise voltage
power $S_{pd}$ at a given current $I$, the tunnel barrier noise
$S_{tb}$ at the same current, and the background noise $S_{th}$ at
zero current. Each of the noise measurements must average more than
1000 noise spectra to give a good signal to noise ratio. From these
averaged noise voltage powers it is possible to extract the Fano
factor of the tunnel barrier at a current $I$. In the limit $eV\gg
k_BT$, where thermal and excess noise can be well separated, the
Fano factor is given by

\begin{equation}
F(I)=\frac{S_{tb}-S_{th}}{S_{pd}-S_{th}}.
\label{equ:fano}\end{equation}

\noindent Each of these calibrated measurements is performed 10-20
times and the results are again averaged. This increases the signal
to noise ratio while minimizing the effects of drifts due to, for
example, temperature changes in the electronics which affect the
gain of the setup. Using this method it is possible to measure the
shot noise from currents smaller than 1 nA, which produce voltage
fluctuations of the order of 1 nV/Hz$^{1/2}$. This is more than 20
times smaller than the thermal noise from the isolation resistors
(Fig. \ref{fig:diagram}), which is the main source of noise in the
system. From now on we will discuss the results of the measurements
of one of our devices, which shows the largest enhancement of the
shot noise.

\section{Results and Discussion}

\begin{figure}[tb]
\begin{center}\leavevmode
\includegraphics[width=0.98\linewidth]{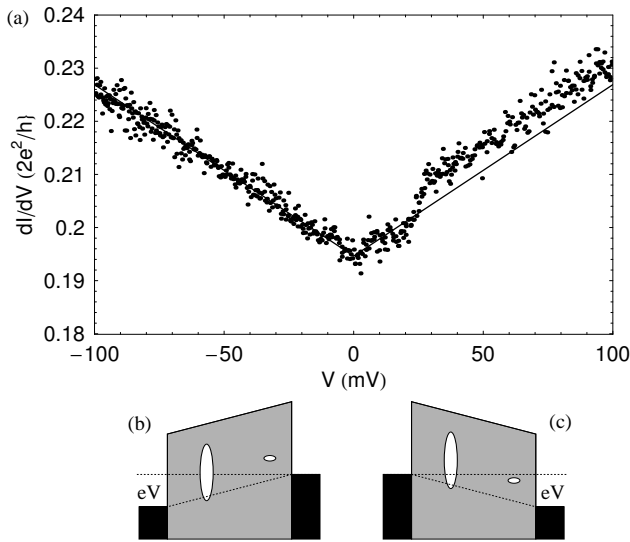}
\caption{(a) Differential conductance in units of $2e^2/h$ measured
by calculating the derivative of resistance data at 8K. The line is
a fit to the negative bias data, and has been symmetrically extended
to the positive bias region. (b,c) Schematics of the asymmetric
configuration of the localized states for negative (b) and positive
(c) bias.}\label{fig:didv}\end{center}\end{figure}

The differential conductance of one of our devices is shown in Fig.
\ref{fig:didv}(a), together with a symmetric linear fit. The most
important features of the differential
conductance are
(i) its asymmetry, (ii) its linearity for negative bias, (iii) the
transition to a higher conductance state above 30mV, and (iv) the
fact that it is non-vanishing at zero bias. The transition to a
higher conductance state which occurs between 20mV and 30mV suggests
that an additional conductance channel, such as a localized state,
with an energy resonance width of the order of 2-3mV, comparable to
$k_BT=0.69mV$, has become available for transport. This localized
state does not become available for transport in the negative bias
regime. This can happen if the position of the localized state
within the barrier is asymmetric as represented by the narrow
resonance localized state in Figs. \ref{fig:didv}(b,c) (small
ellipse towards the right of the barrier). However, even in the
absence of transport through this localized state, the conductance
is nonzero, which means that another channel must be available for
transport at any bias. Furthermore, the conductance of this other
channel increases linearly with the applied voltage. Models for
direct tunneling do not agree with the linear increase in
conductance since they predict a parabolic dependence on voltage.
However, a reasonable scenario where this can happen is tunneling
through an additional localized state above the Fermi energy which
has a very wide resonance [large ellipsoid in Figs.
\ref{fig:didv}(b,c)].

\begin{figure}[tb]
\begin{center}\leavevmode
\includegraphics[width=0.98\linewidth]{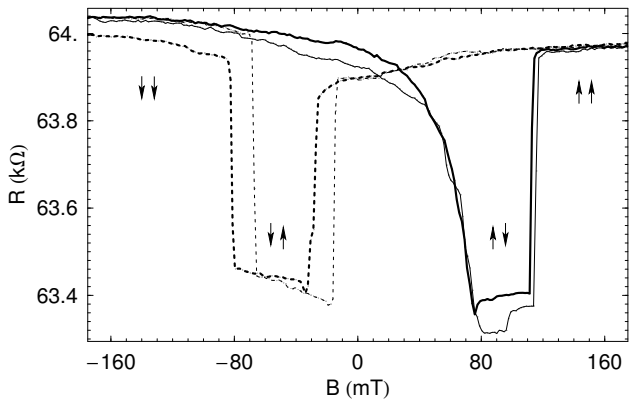}
\caption{Negative magnetoresistance of a Co-Al$_2$O$_3$-Co tunnel
barrier at 8K showing two consecutive
sweeps.}\label{fig:MR8K}\end{center}\end{figure}

The resistance as a function of the in plane field parallel to the
geometrical easy axis of the Co electrodes was then measured at
different biases (Fig. \ref{fig:MR8K} shows the magnetoresistance at
a current of -0.627 $\mu$A). Sharp resistance changes whenever the
magnetization of either of the Co electrodes reverses are evident.
The smoother transition is due to the slower reversal of the wider
Co electrode, in which the effect of a magnetic easy axis not
aligned with the geometrical easy axis is more evident. Two traces
in both directions of magnetic field sweep illustrate that the
transitions can occur at somewhat different fields, showing that the
domain structure configuration after each demagnetization process
can be different. However, there are ranges of magnetic field at
which the device has a definite magnetization configuration and by
applying a magnetic field any of the states $\uparrow\uparrow$,
$\uparrow\downarrow$, $\downarrow\uparrow$, $\downarrow\downarrow$
can be obtained.

An essential observation regarding Fig. \ref{fig:MR8K} is that the
value of the magnetoresistance,
\begin{equation}
MR=\frac{R_{\uparrow\downarrow}-R_{\uparrow\uparrow}}{R_{\uparrow\downarrow}+R_{\uparrow\uparrow}}
\label{equ:MR} \end{equation} is negative. Negative
magnetoresistance was expected and has been observed in
Co-SrTiO$_3$~\cite{deteresa:1999}, Py-Ta$_2$O$_5$, and
Py-Ta$_2$O$_5$/Al$_2$O$_3$~\cite{sharma:1999} systems, and reflects
the importance of the compound effect of spacer and magnetic
materials. However, negative MR has also been observed in systems
which typically exhibit positive MR, and has been attributed to
inversion due to pinhole transport~\cite{mukhopadhyay:026601}, or to
resonant tunneling through a localized state in the tunnel
barrier~\cite{tsymbal:186602}. From the insulator-like temperature
dependence of the resistance we can rule out the possibility of
pinhole conduction. This result, together with the differential
conductance data presented earlier, provide evidence for conduction
through localized states.

\begin{figure}[tb]
\begin{center}\leavevmode
\includegraphics[width=0.98\linewidth]{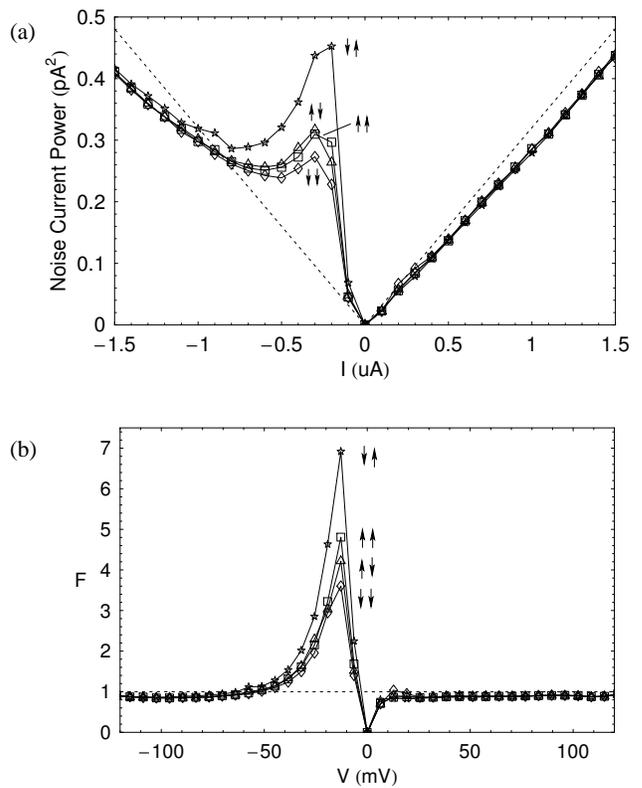}
\caption{(a) Measured shot noise current power as a function of
current at 8K for different magnetization alignments. The dashed
line represents the expected full shot noise. (b) Fano factor
calculated from the data in (a).
}\label{fig:fast}\end{center}\end{figure}

By applying a magnetic field, the sample is set to one of the four
different identifiable magnetization states, $\uparrow\uparrow$,
$\uparrow\downarrow$, $\downarrow\uparrow$, or
$\downarrow\downarrow$, and the shot noise voltage power is measured
as described before. Figure \ref{fig:fast}a shows the results of
such measurements as a function of the current through the device,
together with a dashed line representing the corresponding value of
full shot noise (given by the noise from the photodetector). Figure
\ref{fig:fast}(b) shows the measured Fano factor F as a function of
the voltage across the tunnel barrier, calculated using the data
from Fig. \ref{fig:fast}(a) and Eq. \ref{equ:fano}.

For positive bias the Fano factor is about 0.9, close to the full
shot noise Fano factor of 1. The weak suppression of the shot noise
can happen if electron tunneling through the barrier occurs in a
sequential way through localized states within the barrier or via
parallel channels. For example, tunneling through a localized state
will give a Fano factor between 0.5 and 1 (Eq. \ref{equ:spinblocP}
in the limit $p=0$), depending on the particular values of the
tunneling rates, which themselves depend on the particular bias
conditions, the shape of the tunnel barrier, and the position and
energy of the localized state. These data further supports the claim
for the existence of localized states within the barrier, which we
used to explain the results of the differential conductance and
magnetoresistance measurements. An important observation also
regarding the positive bias region of Fig. \ref{fig:fast}(b) is that
the Fano factor is very similar for the four different magnetization
configurations. This can be explained by Eqs. \ref{equ:spinblocP}
and \ref{equ:spinblocAP} if there exists a large left-right
asymmetry. For example, if $\gamma_L/\gamma_R=6$ and $p=0.1$, the
Fano factor is 0.875 for the P state and 0.877 for the AP state,
which agree with the data within the experimental uncertainty. Even
for higher values of the spin polarization it is possible to find
ratios $\gamma_L/\gamma_R$ which give similar F for both P and AP
states. For example $\gamma_R/\gamma_L=28$ gives 0.877 (0.838) for
the P(AP) for p=0.4. Hence the simple model of tunneling through a
single localized state described in section 3 is compatible with the
positive bias Fano factor behavior.

However, the most interesting result of our measurements comes from
the negative bias part of the shot noise data (Fig. \ref{fig:fast}),
which show a very enhanced voltage dependent shot noise that varies
strongly with the magnetization state of the contacts. As the
voltage bias is decreased from zero the shot noise increases until
it peaks at -13 mV. While the Fano factor for the
$\downarrow\uparrow$ state has a peak value close to 7,
$\downarrow\downarrow$ only increases to 3.6. In addition the state
$\uparrow\downarrow$, which has a similar resistance to
$\downarrow\uparrow$, shows a peak Fano factor of only 4.8. After
-13 mV the shot noise decreases and by -58 mV it is already below
the full shot noise value. At -80 mV the Fano factors for both P and
AP states are close to their positive bias value and, within
experimental uncertainty, they are equivalent. We have already
explained the suppressed positive bias shot noise values as due to
transport through a localized state. In order to understand the
negative bias data we must explain first the existence of peaks in
the shot noise, and second, their spin state dependence.

The simple model of section 3 does not contain any explicit voltage
bias dependence, but it can be included by introducing voltage
dependent tunneling rates, which require assumptions about the
position and energy of the localized state, as well as information
on the properties of the tunnel barrier (in the simplest
approximation the width and height of the barrier). Calculations by
Bulka~\cite{bulka:2000} for magnetic tunnel barriers with a single
localized state have shown that as the voltage bias across the
barrier is increased, the shot noise (after subtracting the thermal
noise) can increase quickly from zero up to a maximum
super-Poissonian value and then decrease with larger bias to
sub-Poissonian values. An asymmetry between positive and negative
bias can be justified for an asymmetric position of the localized
state [Fig. \ref{fig:didv}(b,c)] by observing Fig.
\ref{fig:spinbloc}(a), since for a fixed value of p (say p=0.8),
$\gamma_R/\gamma_L=5$ gives F=0.88 but $\gamma_R/\gamma_L=1/5$ gives
F=3.77. This qualitatively explains the existence of the peaks and
why they appear only for the negative bias region.

However, the large peak values of the Fano factor and their
dependence on the spin state can not be accounted for within the
simple model of transport through a single localized state since it
predicts, on the one hand, a maximum Fano factor of 1.25 for the AP
state and, on the other hand, the same shot noise for both of the P
states (F$_P$=F$_{\uparrow\uparrow}$=F$_{\downarrow\downarrow}$) and
both of the AP states
(F$_{AP}$=F$_{\uparrow\downarrow}$=F$_{\downarrow\uparrow}\neq$F$_P$).
As pointed out by Tserkovnyak (private communication), the
difference in Fano factor between the two P states and between the
two AP states could be due to structural differences between these
states. The magnetoresistance data (Fig. \ref{fig:MR8K}) already
showed additional structure which might suggest that
$\uparrow\uparrow$ and $\downarrow\downarrow$, as well as
$\uparrow\downarrow$ and $\downarrow\uparrow$, are not equivalent
states.
 The large AP
state Fano factor can be explained assuming that two localized
states contribute to transport, as illustrated in Fig.
\ref{fig:spinbloc}(c), where the Fano factor in the AP state can be
not only larger than 1.25 (the maximum value in the case of
transport through a single localized state) but even larger than the
Fano factor for the P state. The dashed lines in Fig.
\ref{fig:spinbloc}(c) illustrate that values of the asymmetry
parameters and the spin polarization (for example, p=0.47) can be
\begin{figure}[bt]
\begin{center}\leavevmode
\includegraphics[width=0.98\linewidth]{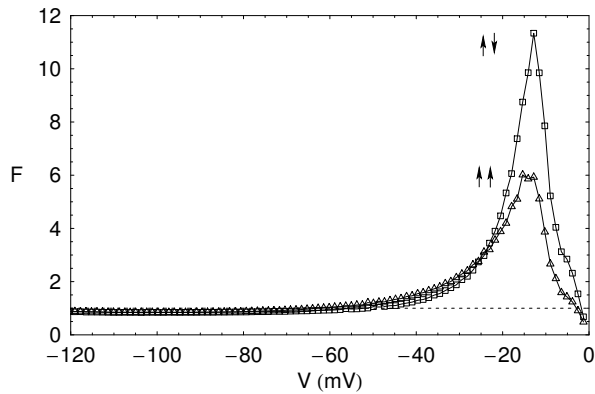}
\caption{Detailed Fano factor curves as a function of voltage for
the same device shown in Fig. \ref{fig:fast} in a different cooldown
to 8K. }\label{fig:fanotwo}\end{center}\end{figure}chosen in order
to give Fano factors which agree ($F_{AP}\simeq 7$, $F_P\simeq 4.7$)
with the measured peak values. Similar Fano factors can be obtained
for both larger and smaller values of the spin polarization by
choosing different combinations of the asymmetry parameters
$\alpha$,$\beta$, and $\gamma$.

Detailed data from a different cooldown in which only
$\uparrow\downarrow$ and $\uparrow\uparrow$ were measured in the
negative bias region are shown in Fig. \ref{fig:fanotwo}. Although
the size of the peaks is different, the voltage at which they occur
is unchanged. The different peak size can be due to differences in
the magnetic states of the electrodes, as well as to differences in
the properties of the localized states after thermal cycling.

\section{Conclusions}

We have made differential conductance, magnetoresistance, and shot
noise measurements in small area magnetic tunnel barriers, which
give evidence for transport through two competing localized states.
The shot noise is enhanced above its Poissonian value at certain
voltages and it is in these regions of super-Poissonian behavior
that a strong dependence of the Fano factor on the magnetization
state of the ferromagnetic electrodes is observed. Such behavior is
explained by using a simple probabilistic model for tunneling
through two strongly interacting localized states, taking into
account the spin dependence of the tunneling rates. The enhancement
of the Fano factor is due to a combination of ``spin blockade" where
minority spin electrons block the transport of majority ones, and
``charge blockade" where an electron that tunnels into a localized
state with small tunneling rate (slow channel) blocks transport
through the other faster channel due to Coulomb interaction. The
voltage dependence of the Fano factor (both the existence of a peak,
and the positive/negative bias asymmetry) can be qualitatively
understood as a change in the tunneling rates when the resonant
energies of the localized states move with respect to the chemical
potentials of each reservoir. However, a complete model which
includes explicit voltage dependence of the conductance and the shot
noise is missing.

We thank Y. Tserkovnyak for helpful discussions.

\bibliographystyle{elsart-num}
\bibliography{DECONS_references}

\begin{thebibliography}{10}
\bibitem{buttcomm}
According to Blanter and Buttiker, classical calculations of the shot noise
  typically agree with the full quantum calculations when ensemble averaging is
  performed. Our simple classical calculations have been compared with those of
  Blanter et. al and Braun et. al and they have been found to agree.
\bibitem{spinflip}
This discussion assumes that there is no spin-flip scattering in the tunneling
  process.
\end{thebibliography}

\end{document}